\documentclass[10pt,conference,letterpaper]{IEEEtran}

\usepackage{times,amsmath,epsfig}
\usepackage{graphicx}
\usepackage{amsfonts}
\usepackage{amssymb}

\newcommand{\Slot}[0]{\mathord{\mathrm{Slot}}}
\newcommand{\Click}[0]{\mathord{\mathrm{Click}}}
\newcommand{\Convn}[0]{\mathord{\mathrm{Purchase}}}

\newcommand{\Choose}[2]{\mathord{{#1} \choose {#2}}}
\newcommand{\Set}[1]{\mathord{\lbrace{#1}\rbrace}}

\newcommand{\eat}[1]{}

\newcommand{\R}[0]{\mathbb{R}}

\def\phi{\varphi}
\def\|{\mathbin{\mid}}

\newtheorem{theorem}{Theorem}

\newtheorem{definition}[theorem]{Definition}

\begin{document}

\title{Toward Expressive and Scalable \\Sponsored Search Auctions}
\author{%
{David J. Martin{\small ${}^{1}$}, Johannes Gehrke{\small ${}^{2}$}, Joseph Y. Halpern{\small ${}^{3}$}}%
\vspace{1.6mm}\\
\fontsize{10}{10}\selectfont\itshape
Department of Computer Science, Cornell University\\
Ithaca, NY, USA\\
\fontsize{9}{9}\selectfont\ttfamily\upshape
${}^{1}$djm@cs.cornell.edu\\
${}^{2}$johannes@cs.cornell.edu\\
${}^{3}$halpern@cs.cornell.edu}

\maketitle
\thispagestyle{empty}

\begin{abstract}
Internet search results are a growing and highly profitable advertising platform.  Search providers auction advertising slots to advertisers on their search result pages.  Due to the high volume of searches and the users' low tolerance for search result latency, it is imperative to resolve these auctions fast.  Current approaches restrict the expressiveness of bids in order to achieve fast winner determination, which is the problem of allocating slots to advertisers so as to maximize the expected revenue given the advertisers' bids.  The goal of our work is to permit more expressive bidding, thus allowing advertisers to achieve complex advertising goals, while still providing fast and scalable techniques for winner determination.
\end{abstract}

\section{Introduction}\label{sec:intro}

With the huge number of Internet searches performed every day, search result pages have become a thriving advertising platform.
The results of a search query are presented to the user as a web page that contains a limited number of slots, typically between four and twenty, for advertisements.
On each search result page, major search engines, like Google and Yahoo, sell these slots to advertisers via an auction mechanism that charges an advertiser only if a user clicks on his ad.
Most of Google's multi-billion dollar revenue, and more than half of
Yahoo's revenue, comes from these so-called sponsored search auctions
\cite{edel05gsp}; and this market is growing quickly.   By 2008,
spending by US firms on sponsored search is expected increase 
by \$3.2 billion from 2006 and will exceed \$9.6 billion, the amount
spent on all of online advertising in 2004 \cite{emar07}. 
With the increasing market size in mind, it is natural to approach sponsored search auctions from a database perspective in order to tackle issues of scalability and expressiveness.
Our paper is a first step in this direction.

Sponsored search auctions currently work as follows:
\begin{enumerate}
\item \textbf{Bid submission.} Advertisers submit bids on clicks for certain keywords offline.
\item \textbf{User search.} A user submits a search query.
\item \textbf{Winner determination.} Slots are assigned to advertisers by the search provider based on the advertisers' bids.
\item \textbf{User action.} The search result page is returned to the user who may now click on one or more of the sponsored links.
\item \textbf{Pricing and payment.} The search provider charges an advertiser according to some pricing rule if the user clicked on the advertiser's sponsored link.
\end{enumerate}

The speed of the winner determination in Step 3 is crucial because it contributes to the user-experienced latency since the winning ads are displayed on the search result page returned to the user.
In current sponsored search auctions, this winner determination can be done quickly because advertisers are limited to submitting a single bid on whether or not the user clicks on their ad.

\subsection{The Need for Expressive Auctions}\label{sec:need}

Unfortunately, as we now point out, the limited bidding in current sponsored search auctions is insufficient to meet advertisers' needs in two respects.

{\bf Bidding on Multiple Features.}
Once the advertisers' ads are displayed on the search results page, the user who submitted the query may click on the ad and may even make a purchase as a result.
Advertisers clearly value purchases because they represent immediate revenue.
They also value clicks on their ads because they indicate potential customers.
However, even if the user does not click on or buy something, advertisers might place value on having their ads displayed simply because this increases their chance to make an impression on the customer.
Advertisers who value brand awareness may wish their ads to be placed in prominent positions.
Such advertisers may prefer their ads to be displayed near the top or bottom of the list, but not in the middle.
Other advertisers whose goals are to be perceived as the leaders in their markets may wish their ads to be displayed in the topmost slot or not displayed at all.
Thus it is clear that advertisers have valuations on clicks, purchases, and slot positions.

Unfortunately, in current search advertising platforms, advertisers are restricted to bidding only on whether they receive a click on their ad.
We call this a \emph{single-feature auction} since the advertisers can express their valuations on only one feature, namely, receiving a click.
Our goal is to support \emph{multi-feature auctions} that would allow advertisers to express valuations on multiple features, namely, clicks, purchases, and slot positions.
Extending bidding to multiple features is non-trivial; 
whereas previously the advertiser submitted a single value as depicted in Figure \ref{fig:tinytbl}, now the advertiser can submit a whole table of values for the different combinations of features, as depicted in Figure \ref{fig:hugetbl}.
The fast algorithms for winner determination that are currently used by Google and Yahoo! do not extend to non-trivial multi-feature auctions.  
Moreover, even for single-feature auctions, these algorithms can correctly deal with only a restricted situation, namely, one where the expected number of clicks on an ad is ``separable'' into the product of an advertiser-specific factor and a slot-specific factor. 

{\bf Dynamic Bidding Strategies.}
The language that search providers, such as Google and  Yahoo, currently use to let advertisers express bidding preferences in is rather limited.
While the language does allow advertisers to specify a limited number of parameters to constrain their bids (such as a daily budget, and geographic targets), the language is often insufficiently expressive for  serious advertisers to express their preferences and how they change over time.  
To deal with this, advertisers employ the services of various third-party search engine management companies (such as iProspect, SureHits, Atlas, etc.) that monitor the outcomes of auctions and periodically resubmit bids on behalf of the advertiser in an attempt to approximate the advertisers' preferences as much as possible.
The kinds of goals that they try to achieve include maintaining a
specified slot position during certain hours of the day, maintaining a
slot position above a specified competitor, and equalizing the return on
investment (ROI) across multiple keywords.%
The success of such search engine management companies demonstrates the
desire among advertisers for more complex expressive bidding in search
auctions. 
Again, advertisers want these, but can only pick from a set of pre-defined strategies that these companies provide. 

\subsection{Our Framework}

With the increasing market size in mind, our goal is to design a framework that allows huge numbers of advertisers to bid on a richer set of features using dynamic bidding strategies while simultaneously allowing the search provider to determine winners quickly so as not to detract from the user experience \cite{bore97latency}.

\textbf{Bidding Language.} 
In this paper, we propose a simple but rich language for bidding that allows advertisers to express their high-level strategies directly; 
we allow users to submit their dynamic strategies as \emph{bidding programs} that can bid on multiple features of the auction outcome, such as purchases and slot positions, in addition to clicks.
Programs take as input the search query and various statistics about auction history and performance, and they output bids on clicks, purchases, and slot positions. 
Using this language gives advertisers direct and fine-grained control over their advertising strategies instead of simply picking from a menu of pre-defined goals, as is currently done. 
Thus, in our framework, the search auctions work as follows:
\begin{enumerate}
\item \textbf{Program submission.} Advertisers submit a bidding program to bid on their behalf.
\item \textbf{User search.}
\item \textbf{Program evaluation.} The programs are run and place bids on clicks, purchases, and slot positions.
\item \textbf{Winner determination.}
\item \textbf{User action.}
\item \textbf{Pricing and payment.}
\end{enumerate}

\textbf{Scalable Algorithms.}
We provide an algorithm for winner determination that takes as
input bids made in our expressive bidding language and runs quickly
provided that the bids satisfy a condition that can be viewed as a
generalization of separability; moreover, we prove that this requirement
is in a sense necessary to get fast performance.

We also provide techniques for reducing the amount of work that needs to
be done when evaluating dynamic strategies of many advertisers.  
This results in a scalable infrastructure for multi-feature auctions
with dynamic strategies.

{\bf Summary of our contributions.} 
We approach sponsored search auctions from a database
perspective, and tackle issues of scalability and expressiveness. 
Our main contribution is an efficient and scalable infrastructure that
permits much more expressive bidding than is currently available. 
In particular, we provide
\begin{itemize}
\item a language to express dynamic bidding strategies for multi-feature
sponsored search auctions (Section \ref{sec:language});
\item an efficient, scalable, and parallelizable algorithm to solve
winner determination for bids in our language (Section \ref{sec:wdp}); 
\item techniques to reduce the amount of work necessary for evaluating
dynamic strategies for multiple advertisers (Section
\ref{sec:multisim}). 
\end{itemize}
We evaluate our techniques experimentally in Section \ref{sec:experiments},
and we conclude in Section \ref{sec:conclusion}.

\section{Bidding Strategies as Programs}\label{sec:language}

In this section, we formalize the notion of bidding on multiple features, and we propose a simple language for dynamic strategies that bid on these features.

\subsection{Multiple Features}\label{sec:multifeature}

Recall that traditionally an advertiser could only bid on one property of the outcome, namely, whether his ad received a click. 
Now we would like to allow advertisers to bid on additional properties as well, namely whether a purchase was made, and whether his ad was displayed within a desired set of slots.
To each advertiser, we make available the following predicates that indicate whether or not the outcome has one of these desired properties.
\begin{enumerate}
\item $\Slot_j$, indicating that the advertiser gets slot $j$, for $j \in \Set{1, \dots, k}$, with $k$ being the number of slots.
\item $\Click$, indicating that the user clicked on the advertiser's ad.
\item $\Convn$, indicating that the user made a purchase via a link from the advertiser's ad.
\end{enumerate}
Conceptually, the advertiser associates a value with each truth assignment to these predicates, as depicted in Figure \ref{fig:hugetbl}.
However, the size of such a representation is exponential in the number of predicates.
So we represent bids as OR-bids on Boolean combinations of predicates instead.
That is, we let the advertiser fill in a \emph{Bids table} where each row corresponds to a Boolean formula of predicates and the amount that he is willing to pay should that formula be true.
If multiple formulas are true, the advertiser can be charged the sum of the corresponding amounts.
For example, the Bids table depicted in Figure \ref{fig:bidtbl} indicates that the advertiser is willing to pay 5 cents if he gets a purchase; 2 cents if his ad is displayed in either positions 1 or 2; and 7 cents if he gets a purchase \emph{and} his ad is displayed in positions 1 or 2.

\begin{figure}[t]
\centering
\begin{tabular}{|c|c|}
\hline
$\Click$ & value \\
\hline
Y & 3 \\
\hline
\end{tabular}
\caption{\normalsize{Single-feature valuation}}\label{fig:tinytbl}
\end{figure}
\begin{figure}[t]
\centering
\begin{tabular}{|c|c|c|c|c|c|}
\hline
$\Convn$ & $\Click$ & $\Slot_1$ & $\Slot_2$ & $\Slot_3$ & value \\
\hline
Y & Y & Y & N & N & 7 \\
N & Y & Y & N & N & 2 \\
\vdots & \vdots & \vdots & \vdots & \vdots & \vdots \\
Y & Y & N & N & Y & 5 \\
N & Y & N & N & Y & 0 \\
\vdots & \vdots & \vdots & \vdots & \vdots & \vdots \\
\hline
\end{tabular}
\caption{\normalsize{Multi-feature valuation}}\label{fig:hugetbl}
\end{figure}
\begin{figure}[t]
\centering
\begin{tabular}{|c|c|}
\hline
formula & value \\
\hline
$\Convn$ & 5 \\
$\Slot_1 \lor \Slot_2$ & 2 \\
\hline
\end{tabular}
\caption{\normalsize{Bids table}}\label{fig:bidtbl}
\end{figure}

\subsection{Dynamic Strategies}\label{sec:programs}

As we said, we are interested in designing a programming language that lets advertisers express more complex preferences, which may change over time.
Instead of providing advertisers with a pre-defined selection of advertising strategies, we let the advertisers submit their bidding strategies as programs for the search provider to run.
Conceptually, each time a user submits a search query to the search provider, these programs are triggered.
The main purpose of these programs is to output bids on clicks, purchases, and slot positions that may result from displaying their ad on the search result page.
In order to do so, each program creates a Bids table as described in Section \ref{sec:multifeature} each time there is a sponsored search auction.
These programs have access to several variables pertinent to the current auction and to the advertiser, such as the keywords in the search query, the time of day, the advertiser's remaining budget, the current return on investment for the keywords that the advertiser is interested in, and so on.
These variables are stored in tables, some of which are read-only shared between all advertisers (such as the time and location of the search) and some of which are private to each advertiser (such as information about the keywords that the advertiser is interested in).
The programs can then be written using simple SQL updates without recursion and side-effects.
SQL triggers can be used to activate programs when an auction begins and to notify programs if they received a slot, click, or purchase.
Programs can modify their private tables, although commonly used variables, such as amount spent, budget remaining, return on investment for various keywords, etc. can be automatically maintained for each program by the search provider.
For example, the advertiser-specific variables related to keywords are stored in a Keyword table, as depicted in Figure \ref{fig:kwdtbl} that is private to each advertiser.
Each tuple in the Keyword table corresponds to a bid for a keyword that the advertiser is interested.  
The attributes of the tuple contain, among other things, the formula for the bid, keyword's relevance score in the search query, the return on investment that this keyword has provided the advertiser, the maximum amount that the advertiser is willing to bid on a click by a user who searched for this keyword, and the amount of money that the advertiser is currently bidding for the keyword.
The search provider updates the return on investment for a keyword each time a user searches for the keyword and then clicks on the advertiser's ad.
The bidding program can be stored with its private tables to improve locality.
Since bidding programs use private tables and read-only shared tables, they do not interact with each other when they are triggered by a new search query.
Hence they can be distributed across several machines and run in parallel if necessary.

\begin{figure}
\centering
\begin{tabular}{|c|c|c|c|c|c|}
\hline
text & formula & maxbid & roi & bid & relevance \\
\hline
boot & $\Click \land \Slot_1$ & 5 & 2 & 4 & 0.8 \\
shoe & $\Click$ & 6 & 1 & 8 & 0.2 \\
\hline
\end{tabular}
\caption{\normalsize{Keywords table}}\label{fig:kwdtbl}
\end{figure}

\subsection{An Example: Equalizing ROI}\label{sec:example}

We now give a concrete example of a dynamic bidding strategy that bids
on multiple features.  Our example combines the dynamic ROI equalizing
heuristic mentioned in Section \ref{sec:intro} with bidding on two
features, clicks and the top slot; the advertiser is interested in
receiving clicks for two keywords, ``boot'' and ``shoe'', but also
wants to be perceived as the leading supplier of boots and so would be
willing to pay extra to be shown in the top slot if the search query
is highly relevant to boots.  In order to control his spending, the
advertiser has a target spending rate that he wishes to maintain.  The
ROI equalizing heuristic, as suggested in \cite{borg07dynamics}, tries
to dynamically allocate spending across the different keywords and
bids so as to maximize the advertiser's ``bang for the buck''.  If the
advertiser is underspending (i.e., his current spending rate is lower
than his target spending rate), then the advertiser increases the bids
on keywords that have been most profitable for him (i.e., those with
the highest return on investment).  If the advertiser is overspending
(i.e., his current spending rate is higher than his target spending
rate), then the advertiser decreases the bids on keywords that have
been least profitable for him (i.e., those with the lowest return on
investment).  Return on investment of a bid is the total value gained
from the keyword (e.g., number of clicks received in the top slot
times the amount the advertiser values a click in the top slot)
divided by the amount spent so far on it.

\begin{figure}
{\small
\begin{verbatim}
 1 CREATE TRIGGER bid AFTER INSERT ON Query
 2 {
 3   IF amtSpent / time < targetSpendRate THEN 
 4     UPDATE Keywords
 5     SET bid = bid + 1
 6     WHERE roi =
 7       ( SELECT MAX( K.roi ) 
 8         FROM Keywords K )
 9       AND relevance > 0 
10       AND bid < maxbid;
11   ELSEIF amtSpent / time < targetSpendRate 
12   THEN
13     UPDATE Keywords
14     SET bid = bid - 1
15     WHERE roi =
16       ( SELECT MIN( K.roi ) 
17         FROM Keywords K )
18       AND relevance > 0 
19       AND bid > 0;
20   ENDIF;
21
22   UPDATE Bids
23   SET value = 
24     ( SELECT SUM( K.bid ) 
25       FROM Keywords K
26       WHERE K.relevance > 0.7
27         AND K.formula = Bids.formula );
28 }
\end{verbatim}}
\caption{\normalsize{Equalize ROI}}
\label{fig:ROI}
\end{figure}

Figure \ref{fig:ROI} shows the program for this strategy.  Line 1 creates
a trigger that waits for a new query to be inserted into the Query
table, indicating that a new auction is taking place.  If the
advertiser notices that he has been underspending (line 3), he
increases his tentative bids for all relevant keywords that have
provided him with the highest ROI, taking care not to increase the bid
past its maximum value (lines 4--10).  Similarly, lines 13--19
decreases his bids for relevant keywords with the lowest ROI if he is
overspending (line 11), taking care not to decrease his bid below zero.
Next, he updates the values in the Bids table
with the sum of his tentative bids for the corresponding formulas for
all sufficiently relevant keywords, namely, those with a relevance
score higher than $0.7$ in the user-submitted search query (lines
22--27).  For example, if the Keywords table is as depicted in Figure
\ref{fig:kwdtbl} after running lines 1--20, then the output Bids table
will be as depicted in Figure \ref{fig:bidtbl2}.

\begin{figure}
\centering
\begin{tabular}{|c|c|}
\hline
formula & value \\
\hline
$\Click \land \Slot_1$ & 4 \\
$\Click$ & 0 \\
\hline
\end{tabular}
\caption{\normalsize{Bids table for Example Program}}\label{fig:bidtbl2}
\end{figure}

\section{Winner Determination}\label{sec:wdp}

Having empowered the advertisers with a language for expressing dynamic bidding strategies to bid on a rich set of features, we now seek efficient and scalable techniques for the search provider to perform winner determination.

All sponsored search auction mechanisms currently in use (see, for example, \cite{agga06truthful,agga06top,edel05gsp,vari05position}) first solve the winner-determination problem, then assign slot positions according to the winning allocation, and finally use some method of charging prices for the positions, such as charging each advertiser their social opportunity cost (this is known as Vickrey pricing \cite{clar71, grov73, vick61sealed}), or charging advertiser in the $k$th slot the amount bid by the next-highest bidder (this is known as generalized second-pricing \cite{edel05gsp}).  
Note that with most pricing schemes, a provider's revenue is \emph{not} the revenue that is computed in the winner-determination problem.  
Different pricing schemes lead to different behavior of the auction in terms of revenue, stability, and other economic and game-theoretic properties.
For example, Vickrey pricing leads to theoretically stable truthful auctions \cite{vick61sealed}, while generalized second pricing leads to locally envy-free equilibria \cite{edel05gsp}.
Nevertheless, the first step in all these auctions is to do winner determination.
Furthermore, given winner determination as a subroutine, the pricing schemes used in these auctions (i.e., Vickrey pricing, generalized second-pricing, etc.) can all be expressed as very simple computations.
In our work, therefore, we focus on optimizing the winner-determination computation.

\subsection{How Winner Determination Works}\label{sec:how-wdp-works}

The \emph{winner-determination problem} is to compute the allocation of slots to advertisers that results in the highest expected revenue for the search engine provider, under the assumption that advertisers actually pay what they bid.  
In keeping with Google and Yahoo policy, we restrict the slot allocations to those in which no advertiser gets assigned more than one slot.
This prevents extremely wealthy advertisers from monopolizing all the available slots.

In order to compute the expected revenue resulting from an allocation, we need the advertisers' bids on clicks, purchases, and slot positions as specified in their Bids tables.
For now, let us assume, that we actually run all of the advertisers' bidding programs to get their resulting Bids tables. 
In Section 4, we give techniques that require us to run only a
small subset of programs under certain conditions. 

In order to compute the expected revenue resulting from an allocation, we also need the probabilities that the formulas in the Bids tables are true in the final outcome.
We thus consider the set of all possible outcomes that describe which slot was allocated to which advertiser together with which advertisers received clicks and purchases.
The probabilities of clicks and purchases depend on the search provider's allocation of slots to advertisers.
For example, ads placed at the top are more likely to be noticed and clicked on than those placed in the middle of the page \cite{niel04branding}.
As a reasonable first-order approximation, we assume that the
probability that a given advertiser gets a click depends only on the
slot allocated to him, and that the probability that he gets a purchase
depends only on whether he got a click and on the slot allocated to
him. 
Furthermore, we assume that the search provider has (or can estimate, using data it has collected) these click and purchase probabilities for each advertiser and each slot allocation to that advertiser.

Note that a complete representation of the probabilities of
all possible formulas for each advertiser is exponential in the number of features. 
Although this is not too large in our setting, the complete set of
probabilities should be stored in a database separate from the run-time
system, which itself should store only probabilities for the formulas
mentioned in the bidding programs and Keyword tables, since these are the
only probabilities that are used. 
Furthermore, the probabilities can be partitioned by advertiser and should be stored with the advertiser's bidding program and private tables to improve locality.

\subsection{Complexity}\label{sec:complexity}

Given the assumptions on slot allocations and distributions above, we look at the complexity of solving the winner-determination problem given bids in our language.
Recall that a bidding program's output is an OR-bid represented by a Bids table whose rows contain bids of the form ``Pay \$$d_1$ for $E_1$'', \ldots, ``Pay \$$d_m$ for $E_m$'', where $E_1, \ldots, E_m$ are Boolean combinations of the $\Slot_j$, $\Click$, and $\Convn$ predicates.
Recall that, in addition, we assume that for any allocation, we have a distribution on outcomes, conditional on that allocation.
Each formula $E_i$ can be identified with an \emph{event} on the set of possible outcomes, namely, the set of outcomes in which $E_i$ is true.
Thus bidding on formulas can be interpreted as bidding on events.
Toward proving that winner determination is tractable for bids in our language, we introduce the following definition. 

\begin{definition}[$m$-dependent event]
An event is \emph{$m$-dependent} if there are at most $m$ advertisers such that probability of the event given any allocation depends only on the placement of those $m$ advertisers.
\end{definition}

That is, an event is $m$-dependent if it is independent of the slots assigned to all but $m$ advertisers.
For example, the event that a given advertiser gets a click is $1$-dependent since we assumed that the probability of an advertiser getting a click depends only on the slot position of that advertiser.
Similarly, the event that a given advertiser is in either the top slot or the bottom slot is $1$-dependent since it depends only on the slot assigned to that advertiser.
However, given two advertisers, the event that one gets the top position and the second gets the bottom is $2$-dependent since it depends on the slots assigned to both those advertisers.

We assume that the representation of each $m$-dependent event includes the labels of the $m$ advertisers on whose slot assignment the event depends.
The following theorem says that winner determination is tractable for $1$-dependent events.%
\footnote{See the Appendix for proofs.} 

\smallskip
\begin{theorem}\label{thm:unary}
For OR-bids on collections of $1$-dependent events, the winner determination problem is in polynomial time.
\end{theorem}
\smallskip

It follows that winner determination for bids
represented by a Bids table can be solved in polynomial time, since 
our assumptions in Section \ref{sec:how-wdp-works} guarantee that any
Boolean combination of predicates for an advertiser (i.e., of the form
$\Slot_1, \ldots, \Slot_k, \Click, \Convn$) is 1-dependent. 

A natural question to ask is whether we can extend our tractability results to a language that allows advertisers to bid on $m$-dependent events, for $m \ge 2$.
The next result says that winner determination is \emph{APX-hard} if we allow bids to be placed on $2$-dependent events, such as the event that one advertiser is displayed above another.
APX is the class of NP optimization problems that have polynomial-time constant-factor approximation algorithms \cite{kann92npo}.  

\smallskip
\begin{theorem}\label{thm:2-dependent}
For OR-bids on collections of $2$-dependent events, the winner-determination problem is APX-hard.  
\end{theorem}
\smallskip

In the remainder of this section, we take the reader on a quest for an efficient and scalable winner-determination algorithm for our bidding language.

\subsection{Existing Allocation Algorithms}

The allocation algorithms used by Google and Yahoo, as well as those studied in the literature \cite{agga06top,agga06truthful,edel05gsp,vari05position}, 
deal with  the issue of scalability by assuming that the probability of a click resulting from assigning a slot to an advertiser is \emph{separable}, that is, it can be written as the product of an advertiser-specific factor and a slot-specific factor. 
To illustrate this notion of separability, we provide examples of
non-separable and separable click probabilities in Figures \ref{fig:nonsep}
and \ref{fig:sep} respectively.  
The matrix in Figure \ref{fig:sep} is separable because the entries in the matrix can be split into the product of advertiser-specific factors (namely, 4 for Nike and 3 for Adidas) and slot specific-factors (namely, 0.2 for slot 1, and 0.1 for slot 2). 

\begin{figure}[htb]
\centering
\begin{tabular}{c|cc}
 & $\Slot_1$ & $\Slot_2$ \\
\hline
Nike    & 0.7 & 0.4 \\
Adidas  & 0.6 & 0.3 \\
\hline
\end{tabular}
\caption{\normalsize{Non-separable click probabilities}}
\label{fig:nonsep}
\end{figure}

\begin{figure}[htb]
\centering
\begin{tabular}{c|cc}
 & $\Slot_1$ & $\Slot_2$ \\
\hline
Nike    & 0.8 & 0.4 \\
Adidas  & 0.6 & 0.3 \\
\hline
\end{tabular}
\caption{\normalsize{Separable click probabilities}}
\label{fig:sep}
\end{figure}

When the click probabilities are separable, it is easy to see that winner determination can be performed by assigning the advertisers with $j$th highest advertiser-specific factor to the slot with the $j$th highest slot-specific factor.  
This can be done in time $O(n \log k)$. 

Note that the assumption of separability implicitly assumes that the event that an advertiser gets a click is $1$-dependent.  
Indeed, it assumes the event that an advertiser gets a click depends on only that advertiser's slot assignment. 
But separability requires much more 1-dependence: it requires that the ratio of the expected number of clicks on one advertiser in a slot and the expected number of clicks on another advertiser in the same slot is the same for all slots.

Not only is separability a much stronger requirement than 1-dependence, but the techniques for fast winner determination that use this assumption do not suffice to deal with our bidding language.  
In particular, they cannot deal with the situations described in Section \ref{sec:intro} where one advertiser wants to be displayed in the top slot or not displayed at all, while another wants to be displayed in either the top or bottom slots but not in the middle slots. 
(Bids representing these preferences can be easily expressed in our language.)

\subsection{Maximum-Weight Bipartite Matching}

We proved Theorem \ref{thm:unary} by showing that winner determination in this case is equivalent to maximum-weight bipartite matching between advertisers and slots, where the edge-weight between an advertiser and a slot is the expected revenue obtained by assigning that slot to that advertiser.
The fastest known (non-parallel) algorithm to solve this is the Hungarian algorithm, invented by Kuhn \cite{kuhn55hungarian} (also known as the Kuhn-Munkres algorithm after being revised by Munkres \cite{munk57assignment}); it finds the best matching in time $O(nk(n + k))$ where $n$ is the number of advertisers and $k$ is the number of slots.
Since this is quadratic in $n$, this will not scale well.
We want to deal with situations where $n$ can be quite large (possibly in tens to hundreds of thousands).  
To make the problem scalable, we need it to be \emph{linear} in $n$, the number of advertisers.  
There are parallel algorithms for maximum-weight matching \cite{fayy06mwbm}, but these require prohibitively large numbers (typically $\Omega(n^2)$) of processing units in order to achieve linear running time.

\subsection{Our Algorithm}\label{sec:algorithm}

\begin{figure*}[t]
\centering
\begin{minipage}{2.1in}
\centering
\vspace{0.3in}
\begin{tabular}{c|cc}
 & $\Slot_1$ & $\Slot_2$ \\
\hline
\textbf{Nike}    & \textbf{9} & 5 \\
\textbf{Adidas}  & \textbf{8} & \textbf{7} \\
\textbf{Reebok}  & 7 & \textbf{6} \\
Sketchers        & 7 & 4 \\
\hline
\end{tabular}
\vspace{0.3in}
\caption{\normalsize{Revenue matrix}}\label{fig:original}
\end{minipage}
\hfill
\begin{minipage}{2.1in}
\centering
\includegraphics[width=2.0in]{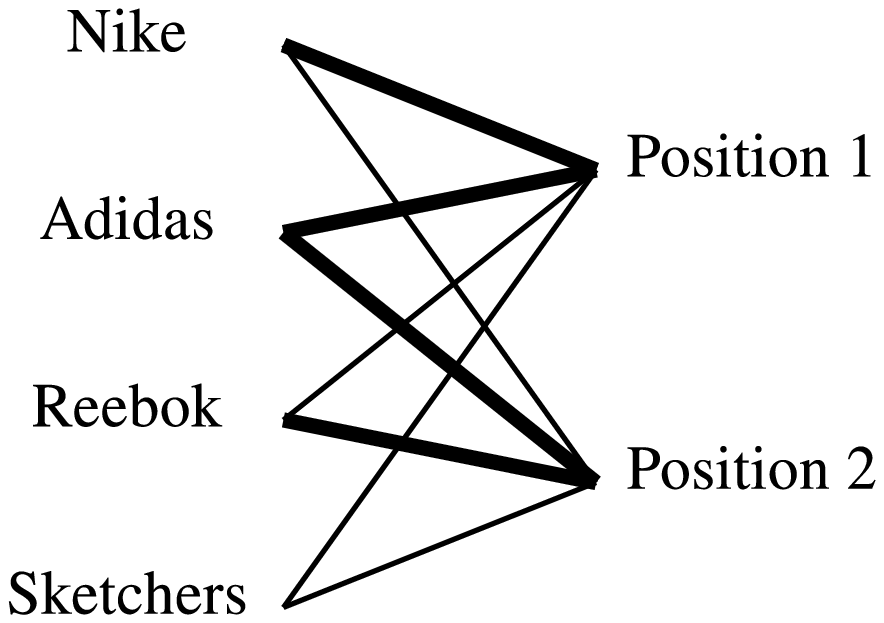}
\caption{\normalsize{Bipartite graph}}\label{fig:bipartite}
\end{minipage}
\hfill
\begin{minipage}{2.1in}
\centering
\vspace{0.3in}
\includegraphics[width=2.0in]{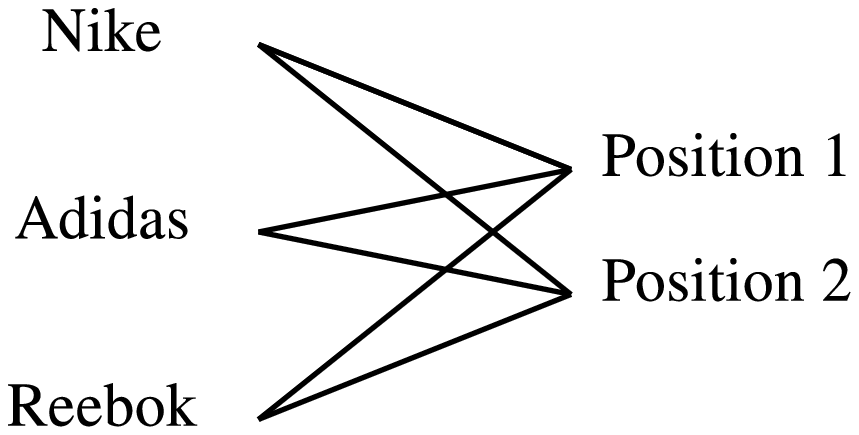}
\vspace{0.11in}
\caption{\normalsize{Reduced graph}}\label{fig:reduced}
\end{minipage}
\end{figure*}

We now give a scalable winner-determination algorithm that takes advantage of the fact that $k$, the number of slots, is quite small (say less than 20) compared to $n$, the number of advertisers.
Indeed, $n$ is growing rapidly every year while $k$ remains the same.
We can modify the Hungarian algorithm to get a $O(n k \log k + k^5)$ algorithm by considering only those advertisers whose values are in the top $k$ highest for some slot.
That is, for each slot, we consider the $k$ advertisers who would produce the top $k$ expected revenue if placed in that slot.
We take the union of these advertisers over all the $k$ slots, and consider the bipartite subgraph containing only these advertisers along with all the $k$ slots.
We then solve maximum-weight bipartite matching problem for this reduced bipartite graph.
As an example, consider the expected revenue matrix as depicted in Figure \ref{fig:original}.
There are two slot positions available and four advertisers.
The top two expected revenues for the first slot come from Nike and Adidas, while the top two expected revenues for the second slot come from Adidas and Reebok.
The corresponding edges in the original bipartite graph between advertisers and slots have been depicted in bold in Figure \ref{fig:bipartite}.
This bipartite graph is then reduced to contain only those advertisers with an adjacent bold edge as depicted in Figure \ref{fig:reduced}.
We observe that the maximum matching for the original problem must occur for this smaller problem since if an maximum matching in the original problem assigned a slot to an advertiser who was not in the top $k$ highest bidders for that slot, we can simply reassign that slot to one of these top $k$ bidders who is not assigned any slot.  Note that since there are only $k - 1$ other slots, at least one advertiser in the top $k$ is guaranteed to remain unassigned.

Finding the relevant advertisers takes time $O(n k \log k)$ because, for each slot, we can find the top $k$ bidders for that slot in time $O(k + n \log k)$ by maintaining a priority heap of size at most $k$.
There are at most $k^2$ such advertisers since in the worst case we will have a distinct set of $k$ advertisers for each of the $k$ slots.
Hence running the Hungarian algorithm on the reduced graph takes time $O(k^5)$ for a total running time of $O(n k \log k + k^5)$ for our algorithm.

\textbf{Parallelization. }
Our technique lends itself very well to parallelization.  Note that in our setting there is typically already a high amount of parallelized infrastructure present since the bids are collected from advertisers in a distributed way.
We construct $k$ networks of computers each in the form of a binary tree of height $O(\log n)$ with $n$ leaves.
We can compute a maximum matching in time $O(k \log n + k^5)$ as follows.
For each slot $j$, we consider the $j$th binary tree network, which will ultimately compute the top $k$ bidders for that slot at the root:

\begin{enumerate}
\item
The $i$th leaf node in the $j$th network starts out with the expected revenue from assigning slot $j$ to advertiser $i$.
\item
Each internal node gathers the top $k$ bidders (along with their corresponding bids) from its two children, and combines them into a single list of top $k$ bidders.  This takes time $O(k)$ for each of the $O(\log n)$ levels of the tree since each level of the tree works in parallel.
\item
The root nodes in each of the $j$-networks take the union of their lists of bidders and compute the maximum-weight matching of these bidders with the $k$ slots using the Hungarian algorithm.
This takes time $O(k^5)$ since there are $k$ slots and at most $k^2$ bidders considered.
\end{enumerate}

Note that we can mix sequential processing with parallel processing by running more than one program sequentially on each machine, computing the top $k$ bids, and then aggregating using a tree network as before.
If we have a binary tree network with $p$ nodes, then the total running time becomes $O(\frac{n}{p} k \log k + k \log p + k^5)$.

Finally the $O(k^5)$ part of the algorithm (i.e., the part resulting
from running the Hungarian algorithm on the reduced bipartite graph) can
be reduced to $O(k^2)$ using a parallel algorithm, such as in
\cite{fayy06mwbm}.  The number of parallel processing units required is
$O(k^5)$, which is independent of $n$. 

\subsection{Beyond 1-dependence}\label{sec:heavyweight}

So far, our results have assumed that the probability that an advertiser receives a click or a purchase depends only on the slot to which that advertiser was assigned.
However, it is easy to think of situations where this assumption might not be true.
For example, if the slot assigned to an advertiser for a small company is just below a very large and popular competitor, then it is likely that the competitor will receive a substantial portion of user clicks that might otherwise have gone to the smaller advertiser had the competitor not been present.
Thus the probability of receiving a click (or a purchase) would depend on who else displays an ad and in what position.
In the worst case, the probability would depend on the entire slot assignment.
The representation of such a general probability distribution would be
quite large ($O(k n^k)$), and, conceptually, winners can be determined
by a brute force algorithm that considers each of the possible
$\Choose{n}{k} k!$ assignments. 

This would also lead to advertisers to value two assignments differently even if both assignments may give the advertiser the same slot.
For example, consider two assignments, both of which assign an advertiser slot 2.
However, in the first assignment, slot 1 is given to a very famous company, while in the second assignment, slot 1 is given to a relatively unknown company.  
Then the advertiser in slot 2 would naturally prefer the second assignment to the first, since the famous company poses a serious threat to the advertiser in terms of diverting away clicks.
Representing such general valuations would also require large space ($O(k n^{k - 1})$) in general.

Motivated by these concerns, but keeping in mind that we cannot store such huge distributions and valuations (since $n$ can be very large), we propose the following model.
For a given search auction, suppose that the advertisers are classified into either \emph{heavyweights} (famous advertisers) or \emph{lightweights} (relatively unknown advertisers).%
\footnote{One way for the search provider to decide which advertisers are heavyweights is to select those advertisers with the most clicks so far.}
We now allow the probability that a given advertiser gets a click (or a purchase) to depend on his slot position as well as on which slots have heavyweight advertisers and which slots have lightweight advertisers.
We also allow advertisers to place bids on which slots get heavyweights and which slots get lightweights, in addition to placing bids on click, purchases, and slot positions as before.
Thus an advertiser might bid 3 cents if he gets slot 2 and if there is a lightweight advertiser in slot 1.
Advertisers could even place more complex bids, such as bidding on having no heavyweights within 3 slot positions above or below his slot in addition to having no more than 2 heavyweights appear anywhere else.
The representation of the probability distributions and valuations now become $O(k 2^{k - 1})$ which does not depend on $n$ anymore.

In order to solve the winner-determination problem, we must find an assignment of slots to advertisers to maximize expected revenue (assuming advertisers pay what they bid) given these new valuations and distributions.
Suppose we knew exactly which slots get heavyweight advertisers in such a revenue maximizing assignment.
We call these slots \emph{heavyweight slots}, and we call the remaining slots \emph{lightweight slots}.
Then we can solve the winner-determination problem by simply solving two disjoint maximum-weighted bipartite bipartite matching problems: one matching the heavyweight advertisers to the heavyweight slots, and the other matching the lightweight advertisers to lightweight slots.
And if we do this for each possible way to choose heavyweight slots, we can find the assignment that maximizes expected revenue over all possible assignments.
Moreover, the maximum-weight bipartite matching problems for different choices of heavyweight slots can be solved independently and in parallel.
Therefore, since there are $2^k$ ways to choose heavyweight slots, we
can solve winner determination in time $O(2^k(n \log k + k^5))$ in
series, or in time $O(n \log k + k^5)$ in parallel using $2^k$
processing units.   
Note that the number of parallel processing units is independent of the
number of advertisers $n$. 

\section{Reducing Program Evaluation}\label{sec:multisim}

We have shown how to solve the winner-determination program given the bids output by programs.  
However, getting these bids for a given search query requires, in the worst case, running each advertiser's program for that query.  
This itself can be quite expensive.
An obvious step toward alleviating this problem is for search providers to use their proprietary keyword matching algorithms to prune away advertisers who are not interested in the search keywords for the current auction.
However, this is not enough if the search query contains a very popular keyword, such as ``music'' or ``book'', where the set of interested advertisers can still be large.
In this section, we show that we can further reduce the amount of work by taking advantage
of knowledge of the structure of the advertiser's programs.
To simplify exposition, we assume that advertisers' programs output bids on only $\Click \land \Slot_1, \dots, \Click \land \Slot_k$.
It is easy to incorporate bids on other formulas since both $\Click$ and $\Convn$ are assumed to be $1$-dependent events.

\subsection{Threshold Algorithm}

We start by considering a situation where the only difference between the programs used by different advertisers is in the values of certain advertiser-specific parameters.  
More precisely, for each slot $j \in [k]$, suppose that each advertiser's bids depends on a set of (numeric) parameters $X_j$ in a monotonic way. 
That is, there is a monotonic function $f_j : X_j \to \R^+$ that takes
as input a value for each parameter in $X_j$ and outputs a bid for a
click in slot $j$. 
We allow some subset of the parameters $Y_j$ to be advertiser-specific: these can vary from advertiser to advertiser (e.g., the amount that they value a particular keyword, the amount of budget remaining, etc.).

Suppose further that these parameters $Y_j$ are updated only by programs
that win the auction.
In Section \ref{sec:logical}, we consider the case where all
programs can update their state; nonetheless, restricting updates to
winning programs is not unreasonable since most useful
advertiser-specific quantities (such as number of auctions won, amount
spent so far, return on investment for a given keyword, etc.) only
change when the advertiser wins an auction.

The rest of the parameters $Z_j = X_j \setminus Y_j$ can be thought of as public global parameters and are the same for all advertisers (e.g., the keyword scores associated with the user's search query, the time and date, the number of times the keywords in search query have appeared today).
A simple example of such a situation is where advertisers all use the same general strategy of starting each day by bidding low and then gradually increasing their bids as the end of the day approaches.  
However, they might each start with a different amount and might increase their bids at different rates.
Then the starting amounts and the rate of increase would be advertiser-specific parameters in $Y_j$, and the time of day would be a global parameter in $Z_j$.

For each advertiser $i$ and each slot $j$, we let the edge weight between advertiser $i$ and slot $j$ be $w_{i, j} \times f_j(y_{i, j}, z_j)$ where $w_{i, j}$ is the probability of advertiser $i$ getting a click in slot $j$, and $y_{i, j} \in Y_j$ are the values of the advertiser-specific parameters and $z_j \in Z_j$ are the values of the global parameters.
We previously showed that we can solve the maximum-weight matching in time $O(n k \log k + k^5)$.
Under the assumptions above, we can further reduce the $O(n k \log k)$
portion that finds the top $k$ bidders for each slot as follows. 
For a given slot $j$, we also store a list of bidders sorted by $w_{i, j}$ and we incrementally maintain $|Y_j|$ lists of bidders, each sorted by one of the parameters in $Y_j$.  
We can then run the \emph{threshold algorithm} \cite{fagi01aggregation} with these lists as input to find the top $k$ advertisers with the highest values of $w_{i, j} \times f_j(y_{i, j}, z_j)$.
Note that we do not need to maintain lists for the parameters in $Z_j$
since all advertisers have the same value for these parameters. 
Since $f_j$ was monotonic, the threshold algorithm is \emph{instance optimal}
for the class of algorithms that find the advertisers with the top $k$
values of $f_j(x_{i, j})$ without making ``wild guesses'' (i.e., the
algorithms must not access an advertiser until that advertiser is
encountered via a sequential scan of one of the lists). 
Instance optimality means that, for any input, the threshold algorithm finds the top $k$ values within a constant factor of the time it takes the fastest algorithm that avoids wild guess on that input.
Given these top $k$ advertisers for each slot, we take $O(k^5)$ further time to compute the winners as described in Section \ref{sec:algorithm}.
To maintain the sorted lists, once the $k$ winners have been computed, we update their $Y_j$ parameters and accordingly update their positions in the sorted lists, which takes $O(|Y_j| k \log n)$ time.

\subsection{Logical Updates}\label{sec:logical}

We now consider the case where all program update their state, not just the winners.
In certain situations, it is possible to reduce the amount of work done in this case as well.
Consider a situation where many programs update their state using an operation that maintains their relative bid ordering.
For example, suppose that many bidders are using the ROI heuristic described in Section \ref{sec:example}, each with possibly different target spending rates and maximum bids.
As long as certain conditions hold (namely, the bid is above zero and the spending rate is above the target spending rate), the heuristic will decrement its bid for a given keyword.
Thus, if we can maintain a \emph{decrement list}---that is, a list of
programs, sorted by their bid, that are currently decrementing their bid
for a given keyword---we can avoid explicitly decrementing each
program's bid, by instead performing a single \emph{logical} decrement
in constant time. 
That is, the decrement list is associated with a single \emph{adjustment variable}, initially zero.
A program's bid is then the sum of the adjustment variable and the program's stored bid.
So, in order to decrement the bids of all programs in the list, we simply decrement the adjustment variable.
The sorted order is maintained because all programs in the list adjust their bids by the same amount.

Of course, the ROI heuristic eventually stops decrementing the bid and starts to increment it (if the spending rate drops below the target) or keep it constant (if the bid is zero) instead.
At this point we must move the program to an \emph{increment list} or a \emph{constant list} as appropriate (similar to a decrement list, except that the adjustment variable respectively increments or remains constant).
At first glance, this would seem to involve checking checking the conditions for each program at every auction.
However, we observe that such conditions can often be reduced to waiting for a shared monotonic variable (such as time, or the number of times a given keyword has occurred) to reach a \emph{critical value}.
For example, in the ROI heuristic, the spending rates of losing programs decreases with time since their amount spent remains constant.  
We can thus compute the next ``critical'' time that a program would have to stop decrementing and start incrementing assuming it continued to lose.
Similarly, we can compute the number of auctions for given keyword necessary before its bid would be decremented to zero and it would have to remain constant at zero.
We maintain a list of \emph{triggers} for the relevant shared monotonic variables, sorted by critical value, that when activated move a bidding program to the appropriate increment, decrement, or constant list, and insert the appropriate new triggers.
This way, we only do work for programs that win an auction and for
triggers whose critical values have been reached.

\section{Experiments}\label{sec:experiments}

\newcommand{\NumSlots}[0]{15}
\newcommand{\NumAuctions}[0]{100}

To evaluate our fast winner-determination algorithm, we compare the
performance of four methods for solving the winner-determination
problem. 
The first method (LP) solves the linear program formulation of the winner-determination problem.
We can prove that this linear program is guaranteed to have an integer
optimum using a theorem of Chv\'{a}tal \cite{chva75}, by showing that the
rows of the constraint matrix represent the maximal cliques of a perfect
graph. 
The second method (H) uses the Hungarian algorithm in a straightforward way to compute the maximum-weight bipartite matching in the bipartite graph with advertisers on the left and slots on the right, where the weight of an edge from an advertiser to a slot is the expected revenue from assigning that slot to that advertiser.
The third method (RH) is our winner-determination technique from Section \ref{sec:algorithm}, which first reduces the bipartite graph.
The fourth method (RHTALU) augments RH with the techniques for reducing program evaluation from Section \ref{sec:multisim} using the threshold algorithm together with logical updates with triggers.

We used $\NumSlots$ slots in all cases.
For simplicity, search queries were generated at a constant rate, each containing one keywords chosen uniformly at random out of 10 keywords.
That chosen keyword was given a relevance score of $1$ for that query, while other keywords had a relevance score of $0$.
All bidders used the ROI heuristic described in Section \ref{sec:programs}.
For each keyword, the bidders' value for a click was generated uniformly at random between 0 and 50
(subject to each bidder having at least one non-zero click value). 
The target spending rates were chosen uniformly at random between 1
and the bidder's maximum value over all keywords. 
The interval $[0.1, 0.9]$ was partitioned into $\NumSlots$ disjoint intervals, with the $(j + 1)$-highest interval associated with slot $j$.
The probability of a given advertiser getting a click in a given slot was generated uniformly at random within that slot's interval.
We used a slight generalization of generalized second-pricing to charge the advertisers who received clicks.

The entire auction system, including the ROI heuristic, was implemented in C++.
We used the GNU Linear Programming Kit to solve the linear program via the simplex method.%
\footnote{We found that the library's interior point method was much slower than the simplex method for our workloads.}
We ran the experiments on an AMD Athlon 64 3800+ processor with 1GB of RAM.

\begin{figure}[t]
\centering
\includegraphics[angle=-90,width=3.2in]{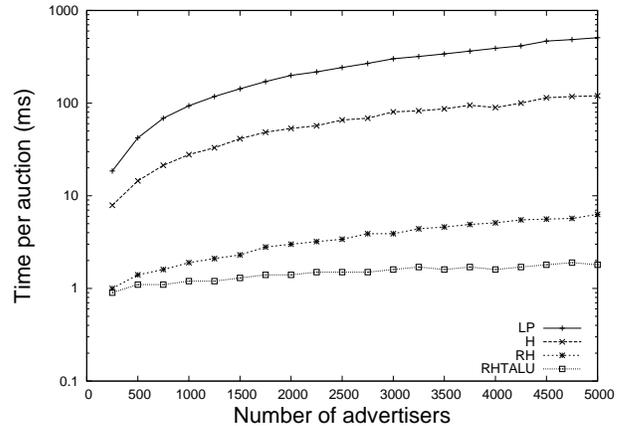}
\caption{\normalsize{Winner Determination Performance}}\label{fig:perf}
\end{figure}

\begin{figure}[t]
\centering
\includegraphics[angle=-90,width=3.2in]{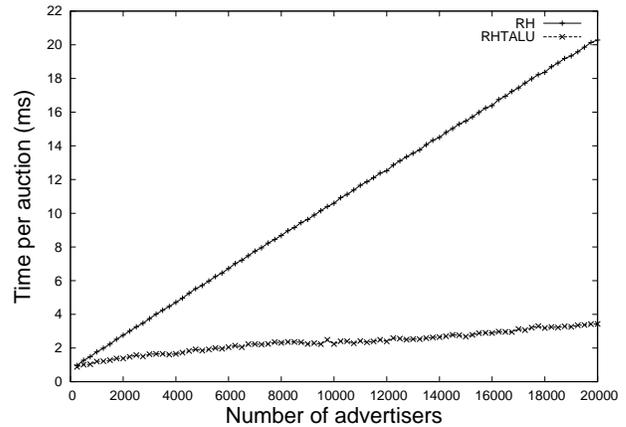}
\caption{\normalsize{Reducing Program Evaluation}}\label{fig:opt}
\end{figure}

Figure \ref{fig:perf} shows, for each of the four methods, the average time taken per auction (over 100 auctions) as we increase the number of bidders.
We observed roughly an order of magnitude improvement of the Hungarian method over naive linear programming solution, and further order of magnitude improvement using our reduced bipartite graph technique.
Figure \ref{fig:opt} compares the performance of methods RH and RHTALU in more detail. It plots the average time taken per auction (over 1000 auctions) as we increase the number of bidders. We observe that our techniques for reducing program evaluation from Section \ref{sec:multisim} give a significant further improvement in performance.

\section{Conclusions}\label{sec:conclusion}

Our paper is a first step toward applying database principles to the exciting and important problems arising in advertising auctions.
In this paper, we highlight the need for more expressive bidding in sponsored search auctions.
To address this, we propose a framework that empowers advertisers with an expressive bidding language, and we provide efficient, scalable, and parallelizable techniques for performing winner determination given bids expressed in our language.

We believe that the database community has much to offer this area given its vast experience with the trade-offs between expressiveness and scalability;
and providing advertisers with more expressive bidding while retaining the scalability of these sponsored search auctions is crucial to the continued growth of this multi-billion dollar industry.

\section*{Acknowledgment}

The authors would like to thank the National Science Foundation (NSF) and the Air Force Office of Scientific Research (AFOSR) for their generous support.
\begin{itemize}
\item
Martin is supported in part by NSF under Grants IIS-0534064 and IIS-0534404. 
\item
Gehrke is in part supported in part by NSF under Grants IIS-0725260 and IIS-0534404, and by AFOSR under Grant FA9550-07-1-0437.
\item
Halpern is in part supported by NSF under Grants ITR-0325453 and IIS-0534064, and by AFOSR under Grant FA9550-05-1-0055. 
\end{itemize}
Any opinions, findings, conclusions or recommendations expressed in this material are those of the authors and do not necessarily reflect the views of the sponsors.

\clearpage
\bibliographystyle{IEEEtran}
\bibliography{IEEEabrv,auctions}
\bigskip
\bigskip
\bigskip
\bigskip
\bigskip
\bigskip
\bigskip
\bigskip
\bigskip
\bigskip
\bigskip
\bigskip
\bigskip

\appendix

\section{Proofs}\label{app:proofs}

\medskip

{\itshape Proof of Theorem \ref{thm:unary}:}
Consider any bid of \$d on event $E$ where $E$ is a $1$-dependent event which depends on the slot assigned to only one advertiser, say $i$.
If advertisers pay what they bid, then in all outcomes this bid contributes exactly the same amount to the revenue as the OR-bid of \$d on $E \land \Slot^i_1$, \$d on $E \land \Slot^i_2$, \dots, \$d on $E \land \Slot^i_k$, and \$d on $E \land (\land_j \lnot \Slot^i_j)$, where $\Slot^i_j$ is the event that advertiser $i$ gets slot $j$.
This is because $\Slot^i_1, \dots, \Slot^i_k$ are mutually exclusive events since the allocations are restricted to at most one slot per advertiser.
We can thus fill out a table of advertisers versus slots where the entry for the $i$th advertiser and the $j$th slot is the sum of the total expected revenue from bids on events of form $E \land \Slot^i_j$ assuming advertisers pay what they bid.
If we interpret this table as the edge-weight matrix of a bipartite graph between advertisers and slots, then the winner-determination problem is the problem of finding a maximum-weight bipartite matching for this graph, which can be done in polynomial time \cite{kuhn55hungarian}.
$\blacksquare$

\bigskip

{\itshape Proof of Theorem \ref{thm:2-dependent}:}
We reduce the winner-determination problem to the maximum-weighted
feedback arc set problem by using bids on $2$-dependent events to encode
the edges in a given weighted directed graphs on advertisers. 
Consider any weighted directed graph on $n$ advertisers.  
Let $w_{i, i'}$ be the weight of the edge from advertiser $i$ to advertiser $i'$.
Let $\Slot^i_j$ be the event that advertiser $i$ gets assigned slot $j$.
For two advertisers $i$ and $i'$, let $E_{i > i'}$ be shorthand for 
$\lor_j (\Slot^i_j \land ((\lor_{j' > j} \Slot^{i'}_{j'}) \lor (\land_{j'} \lnot \Slot^{i'}_{j'}))$, 
which is the event that advertiser $i$ gets a slot and is placed above advertiser $i'$ who may or may not get a slot.
Then $E_{i > i'}$ is a $2$-dependent event since it depends on the slots assigned to advertisers $i$ and $i'$.
Let each advertiser $i$ place the following bids: for each $i' \ne i$, bid $w_{i, i'}$ on $E_{i > i'}$.
Then, assuming advertisers pay what they bid, revenue of $w_{i, i'}$ will be generated if and only if advertiser $i$ is placed above advertiser $i'$.
Then winner determination is equivalent to the problem of finding the maximum-weighted feedback arc set over all size-$k$ subgraphs, which is APX-hard in $n$ and $k$ \cite{kann92npo}.
In fact, even when the directed graphs are restricted to degree-3, the feedback arc set problem is still NP-hard \cite{newm01maximum, karp72reducibility}.
This means that winner determination is NP-hard even when each bid is restricted to the events mentioning at most three advertisers.
So it is infeasible to allow advertisers to bid on being placed above even two or more competitors.
$\blacksquare$

\end{document}